\documentclass[twocolumn,showpacs,preprintnumbers,prl,fleqn]{revtex4}
\usepackage{graphicx}
\def\ul{ }

\begin{document}
\title {Kondo effect in quantum dots coupled to ferromagnetic leads}

\author{J. Martinek,$^{1,2,3}$ Y. Utsumi,$^4$ H. Imamura,$^4$
J. Barna\'s,$^{3,5}$ S. Maekawa,$^2$ J. K\"{o}nig,$^1$  and G.
Sch\"{o}n$^1$}
 \affiliation{$^1$Institut f\"{u}r Theoretische
Festk\"{o}rperphysik, Universit\"{a}t Karlsruhe, 76128 Karlsruhe,
Germany \\
 $^2$Institute for Materials Research, Tohoku University, Sendai
980-8577, Japan \\
 $^3$Institute of Molecular Physics, Polish
Academy of Sciences, 60-179 Pozna\'n, Poland \\
 $^4$Graduate School of Information Sciences, Tohoku University, Sendai
980-8579, Japan \\
 $^5$Department of Physics, Adam Mickiewicz
University, 61-614 Pozna\'n, Poland}

\date{\today}

\begin{abstract}
We study the Kondo effect in a quantum dot which is coupled to
ferromagnetic leads and analyse its properties as a function of
the spin polarization of the leads. Based on a scaling approach we
predict that for parallel alignment of the magnetizations in the
leads the strong-coupling limit of the Kondo effect is reached at
a finite value of the magnetic field. Using an equation-of-motion
technique we study nonlinear transport through the dot. For
parallel alignment the zero-bias anomaly may be split even in the
absence of an external magnetic field. For antiparallel spin
alignment and symmetric coupling, the peak is split only in the
presence of a magnetic field, but shows a characteristic asymmetry
in amplitude and position.
\end{abstract}

\pacs{PACS numbers: 75.20.Hr, 72.15.Qm, 72.25.-b,  73.23.Hk }


\maketitle

The Kondo effect \cite{hewson-book} in electron transport
through a quantum dot (QD) with an odd number of electrons
is experimentally well established \cite{goldhaber,kondo-odd}.
Screening of the dot spin due to the exchange coupling with lead
electrons yields, at low temperatures, a
Kondo resonance. The main goal of the present work is to investigate
how ferromagnetic leads influence the Kondo effect. In the extreme
case of half-metallic leads, minority-spin electrons are
completely absent, i.e., the screening of the dot spin is not
possible, and no Kondo-correlated state can form. What happens,
however, for the generic case of partially spin
polarized leads? How does the spin-asymmetry affect the
Kondo effect? Is there still a strong coupling limit, and how are
transport properties modified?

Based on a poor man's scaling analysis we first show that the
strong-coupling limit can still be reached in this case if an
external magnetic field is applied. This is familiar from the
Kondo effect in QDs with an even number of electrons
\cite{sasaki,cobden,kondo-even-theory-1,kondo-even-theory-2},
which occurs at finite magnetic fields, although the physical
mechanism is different in the present case. In the second part of
the paper we analyze within an equation-of-motion (EOM) approach
the nonlinear transport through the QD. We find that for parallel
alignment of the lead magnetizations the zero-bias anomaly is
split. This splitting can be removed by appropriately tuning the
strength of an external magnetic field $B$. In the antiparallel
configuration of the lead magnetizations no splitting occurs at
zero field.

The Anderson Hamiltonian for a QD with a single level at energy
$\epsilon_0$ coupled to ferromagnetic leads is
\begin{eqnarray}
  H &=& \sum_{rk \sigma}\varepsilon_{rk \sigma}c_{rk\sigma}^{\dagger}
  c_{rk \sigma}
  + \epsilon_0 \sum_{\sigma}d_{\sigma}^{\dagger}d_{\sigma} + U
  d_{\uparrow}^{\dagger} d_{\uparrow} d_{\downarrow}^{\dagger}
  d_{\downarrow}
\nonumber\\
  && + \sum_{rk  \sigma} (V_{rk}  d_{\sigma}^{\dagger} c_{rk \sigma} +
  V^*_{rk} c_{k \sigma}^{\dagger}d_{\sigma} ) +
  g \mu_B B S_{z}
  \; ,
  \label{eq:AMf}
\end{eqnarray}
where $c_{rk\sigma}$ and $d_\sigma$ are the Fermi operators for
electrons with wavevector $k$ and spin $\sigma$ in the leads,
$r=L,R$, and in the QD, $V_{rk}$ is the tunneling amplitude,
$S_{z} = (d_{\uparrow}^{\dagger}
d_{\uparrow}-d_{\downarrow}^{\dagger}d_{\downarrow})/2$, and the
last term is the Zeeman energy of the dot. (Stray fields from the
leads are neglected.) We assume identical leads and symmetric
coupling, $V_{Lk} = V_{Rk}$. The ferromagnetism of the leads is
accounted for by different densities of states (DOS)
$\nu_{r\uparrow}(\omega)$ and $\nu_{r\downarrow}(\omega)$ for up
and down-spin electrons.

In the following we study the two cases of parallel (P)
and antiparallel (AP) alignment of the leads' magnetic moments.
For the AP configuration and zero magnetic field and bias voltage,
the model is equivalent (by canonical transformation
\cite{glazman}) to a QD coupled to a single lead with DOS
$\nu_{L\uparrow}+\nu_{R\uparrow} =
\nu_{L\downarrow}+\nu_{R\downarrow}$. In this case, the usual Kondo
resonance forms, which is the same as for nonmagnetic
electrodes \cite{hewson-book}.

This changes for the P configuration. In this case, there is an
overall asymmetry for up and down spins, say
$\nu_{L\uparrow}+\nu_{R\uparrow} >
\nu_{L\downarrow}+\nu_{R\downarrow}$. To understand how this
asymmetry affects the Kondo physics we apply the poor man's
scaling technique \cite{anderson}, performed in two stages
\cite{haldane}. In the first stage, charge fluctuations dominate
and lead to a renormalization of the QD's levels. Since the
renormalization for the spin-down level is stronger than for
spin-up, a level splitting between the two spin orientations is
generated. This is one of the key mechanism for all the effects
discussed below. To reach the strong-coupling limit it is,
therefore, essential to apply an external magnetic field to
compensate for the generated spin splitting. In the second stage,
the resulting model is mapped onto a Kondo Hamiltonian and the
degrees of freedom involving spin fluctuations are integrated out.
For simplicity we assume for the scaling analysis flat DOSs
$\nu_{r\sigma}$ and neglect the $k$-dependence of the tunnel
amplitudes $V_{rk}=V$.

{\ul First we reduce the cutoff $D$ from $D_0$, which is the smaller
value of the bandwidth and the onsite repulsion $U$ \cite{haldane}}.
Charge fluctuations lead to the scaling equations
\begin{eqnarray}
  \frac{d \epsilon_\sigma }{d \ln( D_0/D)}
  = \frac{ \Gamma_{ \bar{\sigma} } }{ 2 \pi }
\label{eq:Hal_scaling}
\end{eqnarray}
where we defined $\Gamma_\sigma = 2\pi |V|^2 \sum_r
\nu_{r\sigma}$, and $\bar \sigma$ is opposite to $\sigma$. This
yields the solution $\Delta \widetilde{\epsilon} = \widetilde
\epsilon_\uparrow - \widetilde \epsilon_\downarrow = - (1/\pi) P
\Gamma \ln(D_0/D)+ \Delta \epsilon_0 $, where $P = (\nu_{r
\uparrow} - \nu_{r \downarrow})/(\nu_{r \uparrow} + \nu_{r
\downarrow)}$ measures the spin polarization in the leads, $\Gamma
= (\Gamma_\uparrow + \Gamma_\downarrow)/2$ and $\Delta \epsilon_0
= g\mu_B B $ is the Zeeman splitting. The empty-dot state $ | 0
\rangle $ hybridizes with states where the dot is singly occupied
$ |1 \sigma \rangle $ with either spin up or down, while the
singly-occupied state $ |1 \sigma \rangle $  only hybridizes with
the empty-dot state $ | 0 \rangle $ (for $U \gg |\epsilon|$ -
asymmetric Anderson model). Because of the spin-dependent DOS in
the leads the hybridization is spin-dependent, which is the
physical origin of the generated $\Delta \widetilde{\epsilon}$.

To describe Kondo physics (for $\widetilde \epsilon < 0$) we
terminate \cite{haldane} the scaling of Eq.~(\ref{eq:Hal_scaling})
at $\widetilde{D} \sim -\widetilde{\epsilon}$, and perform a
Schrieffer-Wolff transformation.
Using the renormalized parameters $\widetilde{D}$ and $\widetilde{\epsilon}$
we get the effective Kondo Hamiltonian
\begin{eqnarray}
   H_{\rm Kondo} = \sum_{rr'kk'} \{ J_+ S^+c^{\dag}_{rk \downarrow} c_{r'k'
     \uparrow} + J_- S^- c^{\dag}_{rk \uparrow} c_{r'k' \downarrow} +
\nonumber \\
   S_z (J_{z \uparrow} c^{\dag}_{rk \uparrow} c_{r'k' \uparrow} - J_{z
     \downarrow} c^{\dag}_{rk \downarrow} c_{r'k' \downarrow} )  \} \, ,
\label{eq:Kondo}
\end{eqnarray}
plus the term
$-2S_z \tilde D (\nu_\uparrow J_{z \uparrow}-\nu_\downarrow J_{z\downarrow})$
and a potential scattering term.
The initial values for the coupling constants are
$J_+ = J_- = J_{z \uparrow} = J_{z\downarrow} =
{|V|^2 / |\widetilde \epsilon|} \equiv J_0$.
To reach the strong-coupling limit we tune the external magnetic field $B$
such that the total effective Zeeman splitting vanishes, $\Delta
\widetilde{\epsilon} = 0$ {\ul (the field $B$ will also slightly
modify the DOS in the leads \cite{kondo-even-theory-1})}.
During the second stage of scaling spin fluctuations will renormalize the
three coupling constants $J_+ = J_- \equiv J_\pm$, $J_{z \uparrow}$, and
$J_{z \downarrow}$ differently. The scaling equations are
\begin{eqnarray}
   \frac{ d (\nu_\pm J_\pm) }{d \ln(\widetilde D/D) }
   &=& \nu_\pm J_\pm
   (\nu_\uparrow J_{z \uparrow} + \nu_\downarrow J_{z \downarrow} )
\label{eq:scaling1}
\\
   \frac{ d (\nu_\sigma J_{z\sigma}) }{d \ln(\widetilde D/D)} &=&
   2 (\nu_\pm J_\pm)^2
\label{eq:scaling2}
\end{eqnarray}
with $\nu_\pm = \sqrt{ \nu_\uparrow \nu_\downarrow }$, $
\nu_\sigma \equiv \sum_r \nu_{ r \sigma} $ \cite{com_models}. To
solve these equations we observe that $(\nu_\pm J_\pm)^2 -
(\nu_\uparrow J_{z\uparrow}) (\nu_\downarrow J_{z\downarrow}) =0$
and $\nu_\uparrow J_{z\uparrow} - \nu_\downarrow J_{z\downarrow} =
J_0 P (\nu_\uparrow + \nu_\downarrow)$ is constant as well. I.e.,
there is only one independent scaling equation. All coupling
constants reach the stable strong-coupling fixed point $J_\pm =
J_{z\uparrow} = J_{z\downarrow} = \infty$ at the Kondo energy
scale, $D \sim k_B T_K$. For the P configuration the Kondo
temperature in leading order,
\begin{equation}
  T_{\rm K} (P) \approx \widetilde D \exp \left\{ - {1\over
    (\nu_\uparrow + \nu_\downarrow) J_0} \, {{\rm arctanh}(P) \over P}
  \right\} \; ,
\label{eq:Kondo_temperature}
\end{equation}
depends on the polarization $P$ in the leads.
It is a maximum for nonmagnetic leads, $P=0$,
and vanishes for $P \rightarrow 1$.

Finally, we point out an interesting consequence of the spin
polarization in the leads. With nonmagnetic leads, the Kondo
Hamiltonian couples the spin of the QD to the spin of the leads
only, but not to its charge. To analyze the analogous situation in
our case, we introduce the (pseudo) spin $\vec \sigma = (1/2)
\sum_{kk'rr'\sigma\sigma'} c^\dagger_{kr\sigma} \sigma_{\sigma
\sigma'} c_{k'r'\sigma'} / (2 \widetilde{D}\sqrt{\nu_\sigma
\nu_{\sigma'}})$, where the spin-dependent normalization factor is
crucial to ensure the proper spin commutation relations, and the
(pseudo) charge $en = e\sum_{kk'rr'\sigma} c^\dagger_{kr\sigma}
c_{k'r'\sigma} / (2 \widetilde{D} \nu_\sigma)$. The last term in
Eq.~(\ref{eq:Kondo}) can, then, be written as $2\widetilde{D}
(\nu_\uparrow J_{z \uparrow} + \nu_\downarrow J_{z \downarrow})
\sigma_z S_z$ plus $\widetilde{D} (\nu_\uparrow J_{z \uparrow} -
\nu_\downarrow J_{z \downarrow}) n S_z$. The first term is
analogous to the Kondo model with nonmagnetic leads, while the
second term couples spin to charge.
The latter does not scale up
and the associated additional renormalization of the Zeeman
splitting, $-(1/\pi)P\Gamma$, is negligible as compared to $\Delta
\widetilde{\epsilon}$ in the limit $ D_0 \gg |\epsilon| $.

The unitary limit for the P configuration can be achieved by
tuning the magnetic field appropriately, as discussed above. In
this case, the maximum conductance through the QD is $G_{{\rm
max},\sigma}^P = e^2/h$ per spin, i.e., the same as for
nonmagnetic leads. This yields that the amplitude of the Kondo
resonance for up and down spin at the Fermi level are different,
since $G_{{\rm max},\sigma}^P \sim \Gamma_\sigma(E_{\rm F})
\rho_\sigma(E_{\rm F})$, and therefore, $\rho_\uparrow(E_{\rm
F})/\rho_\downarrow(E_{\rm F}) = (1-P)/(1+P)$. For the AP
configuration, the maximal conductance is reduced, $G_{{\rm
max},\sigma}^{AP} = (1-P^2) e^2/h$, and vanishes for $P
\rightarrow 1$.

In the remainder of this article we analyze the DOS of the QD and
address nonequilibrium transport. For a qualitative discussion we
should employ the simplest technique which accounts for both the
formation of Kondo resonances and the influence of the
spin-dependent renormalization of the dot level on spin
fluctuations. The equation-of-motion (EOM) technique with the
usual decoupling procedure \cite{meir,EOM} for higher order Green
functions (GF) satisfies the first requirement but not the second.
We, therefore, extend this scheme by calculating the level
splitting $\Delta \widetilde{\epsilon}$ self-consistently. For a
more quantitative analysis one could include higher-order (than
usually) Green's functions in the EOM approach or higher-order
diagrams in the resonant-tunneling-approximation \cite{konig}, or
use more advanced schemes such as real-time \cite{konig2} or
numerical RG \cite{costi1} methods. These techniques are, however,
much more complex \cite{com_slaveboson} and are not pursued here.

Within the Keldysh formalism, the transport current $ I =
\sum_{\sigma} I_{\sigma} $ through a QD for $ \Gamma_{{\rm R}
\sigma }(\omega) = \lambda_{\sigma} \Gamma_{{\rm L} \sigma
}(\omega) $ is
\begin{eqnarray}\label{eq:Current}
I_{\sigma}=\frac{e}{\hbar} \int d\omega \frac{ \Gamma_{{\rm L}
\sigma}(\omega) \Gamma_{{\rm R} \sigma }(\omega) }{ \Gamma_{{\rm
L} \sigma}(\omega) + \Gamma_{{\rm R} \sigma }(\omega)  } [f_{\rm
L}(\omega)-f_{\rm R}(\omega)] \rho_{\sigma} (\omega)
\end{eqnarray}
where $\rho_\sigma (\omega ) = -(1/\pi) \, {\rm Im} \, G^{\rm
ret}_\sigma(\omega) $. For strong interaction ($U\rightarrow
\infty $) the retarded Green's function is
\begin{eqnarray}
\label{eq:Green_f}
   G^{\rm ret}_\sigma(\omega) = \frac{1-\langle
     n_{\bar{\sigma}} \rangle}{ \omega -\epsilon_{\sigma}
     -\Sigma_{0\sigma}(\omega)-\Sigma_{1\sigma}(\omega) + i0^+}
\end{eqnarray}
where $\Sigma_{0\sigma}(\omega)  = \sum_{k \in  {\rm L,R}} | V_{k} |^2 /
( \omega -\varepsilon_{k \sigma} )$ is the self-energy for a noninteracting QD,
while
\begin{eqnarray}
\label{eq:Self_energy}
   \Sigma_{1\sigma}(\omega, \Delta \widetilde{\epsilon} ) =
   \sum_{k \in  {\rm L,R}}  \frac{| V_{k} |^2 f_{\rm L/R}
     (\varepsilon_{k \bar{\sigma} } ) }
   { \omega - \sigma \Delta \widetilde{\epsilon}
     -\varepsilon_{k \bar{\sigma} } + i \hbar /2
     \tau_{ \bar{\sigma} }  }
\end{eqnarray}
appears for interacting QDs only.
The average occupation of the QD with spin $\sigma$ is obtained from
$\langle n_{\sigma} \rangle = -( i/2 \pi ) \int d \omega \,
G^<_\sigma(\omega)$.
Extending the standard derivation \cite{meir}, we replaced on the
r.h.s.\ of Eq.~(\ref{eq:Self_energy}) $\Delta \epsilon \rightarrow
\Delta \widetilde{\epsilon}$, where $\widetilde{\epsilon}_\sigma$
is found self-consistently from the relation
\begin{eqnarray}
\label{eq:Self_consist}
   \widetilde{\epsilon}_{ \sigma } = \epsilon_{\sigma} + {\rm Re}
   [ \Sigma_{0\sigma}(\widetilde{\epsilon}_{ \sigma }) +
   \Sigma_{1\sigma}(\widetilde{\epsilon}_{\sigma } , \Delta
   \widetilde{\epsilon} ) ] \, ,
\end{eqnarray}
{\ul which describes the renormalized dot-level energy, where the
real part of the denominator of Eq.~(\ref{eq:Green_f}) vanishes
\cite{hewson-book}.} We emphasize that without this
self-consistency relation the Kondo resonances will, in general,
appear at different positions, which disagree with the conclusions
from the scaling analysis \cite{sergueev}.
 The procedure simulates
higher-order contributions and the influence of the
renormalization of the dot level on spin fluctuations. Following
Ref.\ \cite{meir} we introduce, in heuristic way, a lifetime
$\tau_{\sigma}(\mu_{ \rm L}, \mu_{ \rm R}, \widetilde{\epsilon}_{
\uparrow }, \widetilde{\epsilon}_{\downarrow } )$ which describes
decoherence due to a finite bias voltage $V$ or level splitting
$\Delta \widetilde \epsilon$. It is obtained in second-order
perturbation theory and depends on the electrochemical potentials
in the leads, $\mu_{\rm L(R)} $, and the level positions. Again,
we replace the bare levels by the renormalized ones. In the
numerical results presented below we use Lorentzian bands of width
$D = 100 \Gamma$.

For nonmagnetic leads, $P=0$ and zero magnetic field, $B=0$, the proposed
approximation is identical to the standard EOM scheme \cite{meir}.
For finite magnetic field, $B\ne0$, the self-consistency condition yields a
splitting of the Kondo resonances which is slightly smaller than $2g\mu_B B$,
in agreement with both experimental \cite{goldhaber} and theoretical
findings \cite{moore,costi2}.
For $B = 0$ and $P > 0$ in the parallel configuration, we obtain a value
of the splitting $\Delta \widetilde{\epsilon}$ comparable to the result from
scaling, Eq.~(\ref{eq:Hal_scaling}).

\begin{figure}[h!]
\centerline{\includegraphics[width=0.8\linewidth]{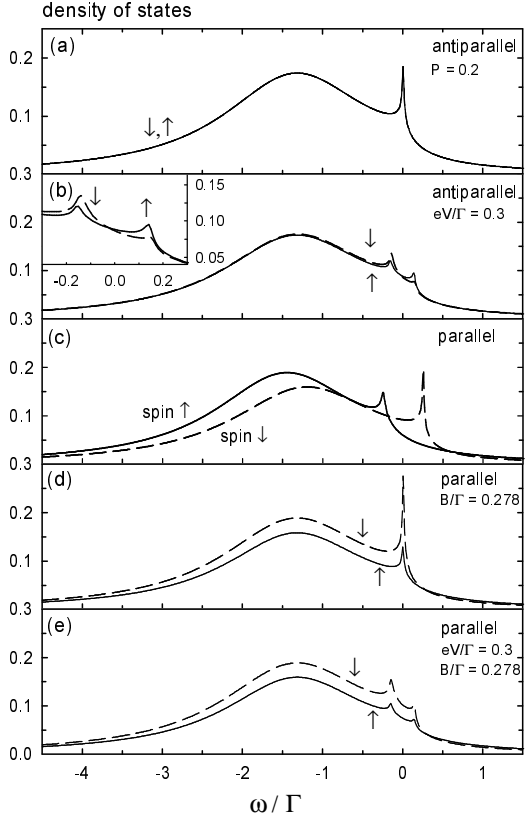} }
 \caption{Spin
dependent DOS for spin up (solid line) and spin down (dashed),
calculated for P and AP alignment (as indicated), for a spin
polarization of the leads $ P = 0.2 $. The parts (d,e) include the
effect of an applied magnetic field $B$ and (b,e) of an applied
bias voltage $V$. The other parameters are: $ T/\Gamma = 0.005 $
and $ \epsilon /\Gamma = -2 $.} \label{figure1}
\end{figure}

In Fig.~\ref{figure1} we plot the DOS of the QD for spins in AP
and P configurations with spin polarization
$P=0.2$ in the leads. In the AP configuration there is one Kondo
resonance [Fig.~\ref{figure1}(a)] and the DOS is the same as for the
case of nonmagnetic leads. For the P configuration, however, the
Kondo resonance splits [Fig.~\ref{figure1}(c)], which can be
compensated by an external magnetic field $B$
[Fig.~\ref{figure1}(d)]. In the latter case, the amplitude of the
Kondo resonance for spin down significantly exceeds that for
spin up (as discussed above). A finite bias voltage,
$\mu_{\rm R} - \mu_{\rm L} = eV>0$, again leads to a splitting
for both the AP and the P configurations,
[Fig.~\ref{figure1}(b) and (e)]. In the AP configuration, the
amplitude of the upper and the lower Kondo peak appear asymmetric
[Fig.~\ref{figure1}(b)].

\begin{figure}[h!]
\centerline{
\includegraphics[width=1\linewidth]{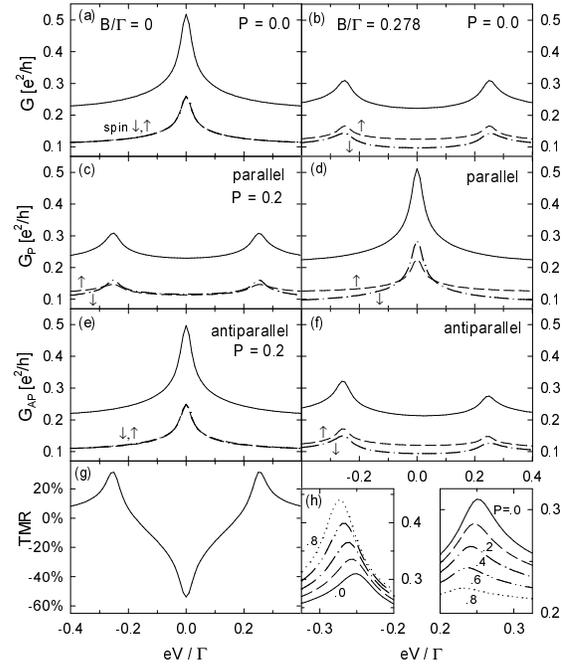} }
 \caption{Total
differential conductance (solid lines) as well as the
contributions for spin up (dashed) and the spin down
(dotted-dashed) vs. the applied bias voltage $V$ at zero magnetic
field $ B = 0 $ (a,c,e) and at finite magnetic field (b,d,f,h) for
normal (a,b) and ferromagnetic leads with parallel (c,d) and
antiparallel (e,f,h) alignment of the lead magnetizations. Part
(g) shows the tunnel magnetoresistance, $ { \rm TMR } = (G_{\rm
P}-G_{\rm AP})/G_{\rm AP} $, for the cases (c) and (e). (h)~The
conductance $G_{\rm AP}/(1-P^2)$ for several values of $P$ as
indicated. Other parameters are as in Fig.~\ref{figure1}}
 \label{figure2}
\end{figure}

In Fig.~\ref{figure2} we show the differential conductance as a
function of the transport voltage. For nonmagnetic leads, there is
a pronounced zero-bias maximum [Fig.~\ref{figure2}(a)], which
splits in the presence of a magnetic field [Fig.~\ref{figure2}(b)]. For
magnetic leads and parallel alignment, we find a splitting of the
peak in the absence of a magnetic field [Fig.~\ref{figure2}(c)],
which can be tuned away by an appropriate magnetic field
[Fig.~\ref{figure2}(d)]. In the AP configuration, the opposite
happens, no splitting at $B=0$ [Fig.~\ref{figure2}(e)] but finite
splitting at $B>0$ [Fig.~\ref{figure2}(f)] with an additional
asymmetry in the peak amplitudes as function of the bias voltage.
This asymmetry is related to the asymmetry in the
amplitude of DOS [Fig.~\ref{figure1}(b)].
 All these findings are
in good agreement with our scaling analysis. In
Fig.~\ref{figure2}(g) we show the tunnel magnetoresistance (TMR)
which can be much larger than for conventional TMR systems.
Finally, we find that the positions of the peaks in the AP
configuration in the presence of a magnetic field are slightly
shifted as a function of the polarization $P$
[Fig.~\ref{figure2}(h)]. This can be explained in
the similar way as in Ref.~\cite{kondo-even-theory-2} by an
additional level splitting $\delta \Delta \widetilde{\epsilon} =
(1/4 \pi) \, P \, \Gamma/\widetilde{\epsilon} \, e V$ at finite
bias voltages due to spin accumulation in the QD.


We finally comment on the observability of our proposal. Ferromagnetic
systems usually have strong spin-orbit coupling which could lead to
spin-flip relaxation and, thus, could destroy the Kondo effect. On
the other hand, recent observation of Aharonov-Bohm oscillations
in a ferromagnetic ring \cite{kasai} proved coherent transport,
i.e., the formation of the Kondo cloud in ferromagnetic leads might
be possible as well. How can one attach ferromagnetic leads to a QD?
A conceivable realization might be to put
carbon nanotubes in contact to ferromagnetic leads \cite{ago}. The Kondo
effect has been observed already  for nonmagnetic electrodes
\cite{cobden,bachtold}. Alternatively one  might
use magnetic tunnel junctions with magnetic impurities in the
barrier, or spin-polarized STM \cite{ralph,wiesendanger}.

In conclusion, we presented a qualitative study of the Kondo effect
in QDs coupled to ferromagnetic leads.
In particular we found a splitting of the Kondo resonance for parallel
alignment of the leads magnetizations, even in the absence of a magnetic
field.
Our results are based on a poor man's scaling approach and an EOM technique.
Further investigations on a more quantitative level using more advanced
techniques would be desirable.

We thank L. Borda, J. von Delft, {\ul L. Glazman, B. Jones,} Yu.V.
Nazarov, A. Rosch, H. Schoeller, A. Tagliacozzo, M. Vojta and
A.Zawadowski for discussion. Support by the Research Project KBN 5
P03B 091 20 and the DFG through the CFN and the Emmy-Noether
program is acknowledged.



\begin{references}
\bibitem{com_models}
We note that the (spin-dependent) DOS
$\nu_\sigma$ and the (spin-independent) tunneling amplitudes $V$ enter
the scaling equations Eqs.~(\ref{eq:Hal_scaling}), (\ref{eq:scaling1}) and
(\ref{eq:scaling2}) only via the combination $\nu_\sigma |V|^2$.
As a consequence, all the conclusions drawn in this paper are valid
also for a system with spin-dependent tunneling amplitudes but nonmagnetic
leads.
\bibitem{EOM}
The approximations used here within the EOM scheme are rather
crude. Its validity range is limited to temperatures above $T_{\rm
K}$. In this regime, while quantitative precision for quantities
such as the Kondo temperature and the occupation number are not
achieved the main physical effects are well described (see
Ref.~\cite{meir} and references therein).
\bibitem{com_slaveboson}
The slave boson technique in the non-crossing approximation
does not describe the influence of magnetic fields properly
\cite{meir}, whereas the mean field approximation
overestimates the magnetic order, especially \cite{dorin} for
large $ U $, since it neglects quantum fluctuations.
\end{references}
\end{document}